\documentclass[10pt]{iopart}

\usepackage{iopams}  
\usepackage[numbers]{natbib}  
\usepackage{graphicx,color}
\usepackage{braket}

\usepackage[colorlinks={true}, citecolor={blue}, filecolor={blue}, linkcolor={blue}, urlcolor={blue}]{hyperref}

\usepackage{xcolor}

\begin{document}

\title{Spin Squeezing Enhanced Quantum Magnetometry with Nitrogen-Vacancy Center Qutrits}
\author{Lea Gassab\footnote{Author to whom any correspondence should be addressed.}}
\ead{leagassab974@gmail.com}
\address{Department of Physics, Ko{\c{c}} University, 34450 Sar{\i}yer, Istanbul, T\"{u}rkiye}
\address{Departments of Biology, Physics \& Astronomy, Waterloo Institute for Nanotechnology, University of Waterloo, Waterloo ON Canada}
\author{\"{O}zg\"{u}r E. M\"{u}stecapl{\i}o\u{g}lu}
\address{Department of Physics, Ko{\c{c}} University, 34450 Sar{\i}yer, Istanbul, T\"{u}rkiye}
\address{T\"{U}B\.{I}TAK Research Institute for Fundamental Sciences, 41470 Gebze, T\"{u}rkiye}
\address{Faculty of Engineering and Natural Sciences, Sabanc{\i} University, 34956 Tuzla, Istanbul, T\"urk{\.i}ye}


\begin{abstract}
We explore the utility of quantum spin squeezing in quantum magnetometry, focusing on three-level (qutrit) Nitrogen-Vacancy (NV) centers within diamond, utilizing a standard Ramsey interferometry pulse protocol. Our investigation incorporates the effects of dephasing and relaxation on NV centers' dynamics during Ramsey measurements, modeled via the Lindblad quantum master equation. We conduct a comparative analysis between the metrological capabilities of a single NV center and a pair of NV centers, considering Quantum Fisher Information both with and without spin squeezing. The quantum correlations between NV centers are assessed through the evaluation of the Kitagawa-Ueda spin squeezing parameter within a two-level manifold. Additionally, parallel calculations are conducted using a two-level model (qubit) for NV centers. Our findings reveal that leveraging qutrits and spin squeezing yields enhanced magnetometric precision, albeit constrained by dephasing effects. Nevertheless, even in the absence of dynamical decoupling methods to mitigate environmental dissipation, strategic timing of squeezing and free evolution can sustain the advantages of qutrit-based magnetometry.
\end{abstract}

%
%
%
%
\ioptwocol

\section{Introduction}
The exploration of precisely controlled quantum systems as highly sensitive nanoscale detectors holds significant promise in advancing our comprehension of intricate processes within biological and condensed-matter systems, operating at molecular and atomic scales \cite{hasselbach2000microsquid, kirtley1995high}. The demanding criteria for both high sensitivity and spatial resolution have prompted suggestions to employ spin-based quantum systems as nanoscale magnetometers \cite{chernobrod2005spin} or to utilize imaging through the detection of sample-induced decoherence~\cite{cole2009scanning}. A particularly appealing physical platform for realizing these concepts is the Nitrogen-Vacancy (NV) center in diamond. This choice is motivated by the center's prolonged coherence times at room temperature and its convenient optical readout of the spin state \cite{degen2017quantum,jelezko2006single,balasubramanian2009ultralong,hanks2017high,hong2013nanoscale,schirhagl2014nitrogen}.

Given the considerable potential of the NV center, researchers have conducted studies to enhance the sensing capabilities of this probe. Specific pulse sequences, such as Carr-Purcell-Meiboom-Gill (CPMG) and Dynamic Decoupling (DD) \cite{shim2012robust,wang2012comparison}, have been used to prolong coherence time. Additionally, researchers have investigated the use of entangled NV centers \cite{qiu2021nuclear,xie2021beating} and, finally, spin squeezing \cite{dooley2016hybrid}. Spin squeezing, as introduced by Kitagawa and Ueda \cite{kitagawa1993squeezed}, involves the quantum redistribution of uncertainties along two orthogonal spin directions. This operation holds significance in quantum sensing and metrology, providing a means to enhance measurement precision beyond the standard quantum limit in experiments \cite{brask2015improved,buchmann2016complex,gross2012spin}. Consequently, spin squeezing emerges as a valuable resource in applications of quantum technology. In the context of spin squeezing generation for NV centers, ensembles of qubits are typically considered. However, this approach has limitations as the qubit model is an approximation that necessitates a bias field. Here, we explore the potential of using qutrits for NV centers. Pulsed dynamical generation of spin squeezing has been explored in the literature for spin-1 systems~\cite{begzjav2021squeezing,huang2021dynamic} and other methods, including generation via nonadiabatic control~\cite{xin2022fast}. In the context of NV centers, Ramsey interferometry is recognized as a straightforward and efficient protocol for measuring magnetic fields. It has been extensively studied in qubit systems \cite{guldeste2023wavelet,oshnik2022robust,coto2021probabilistic,rondin2014magnetometry} and has also been discussed for qutrit systems \cite{oon2022ramsey,hart2021n}.

In this article, we explore the potential enhancements in quantum metrological performance achievable by transitioning from two-level (qubits) to three-level (qutrits) spin systems and incorporating spin squeezing within typical quantum magnetometry settings, specifically NV center Ramsey interferometry. We adopt a foundational approach, comparing qubit versus qutrit Ramsey quantum magnetometry scenarios, both with and without spin squeezing, in terms of Quantum Fisher Information (QFI). Considering the influence of quantum dephasing and relaxation during the spin system's evolution, our findings suggest that qutrits, coupled with the additional step of spin squeezing, can outperform qubits in terms of QFI, provided that the spin squeezing magnitude, the timing of squeezing and free evolution are optimally determined. All results presented herein are derived from numerical simulations, which offer quantitative insight into the complex dynamics and metrological performance of these quantum systems.

The structure of the paper is as follows: Section~\ref{sec:model} provides an overview of the employed model, elucidating the Hamiltonian and dynamics. In Section~\ref{fisher}, we introduce QFI and spin squeezing parameter. Section~\ref{results} presents our results, and finally, Section~\ref{conclusion} summarizes our conclusions.

\section{Model}
\label{sec:model}

In a scenario where an unknown magnetic field is to be estimated, we consider a sensing protocol based on NV centers that are initialized and controlled in the presence of the same field. Although the field strength is not precisely known in advance, we assume it to exceed a few Gauss—sufficient to lift the degeneracy between the \(\ket{+1}\) and \(\ket{-1}\) spin states via Zeeman splitting. This splitting, typically on the order of several MHz or more, enables frequency-selective microwave control. In particular, the \(\ket{0} \leftrightarrow \ket{-1}\) transition can be addressed resonantly without significantly exciting \(\ket{+1}\), since typical microwave pulses have bandwidths of only a few MHz~\cite{barry2020sensitivity}. To accomplish this, we utilize one or two NV centers as our probes. The NV center, characterized as an optically active color defect, results from a substitutional nitrogen impurity and a neighboring carbon vacancy in the diamond lattice. Comprising six electrons from nitrogen and the surrounding carbon atoms, the negatively charged NV center is described as a spin-1 system. We opt to explore the full capabilities of the NV center by modeling it as a qutrit system. This choice is motivated by the potential advantages that qutrits offer in metrology when compared to qubits~\cite{yan2010optimal,shlyakhov2018quantum}. Its ground state is a spin triplet ($^3A_2$) denoted as $\ket{Sm_S}$ with $S = 1$ and $m_S = 0, \pm 1$. The NV center Hamiltonian captures these properties, while the excited-state triplet ($^3E$) is acknowledged at a higher energy level and is not explored in this context~\cite{barry2020sensitivity,doherty2013nitrogen}.

\begin{figure}[t!]
		\centering
  		\includegraphics[width=\linewidth]{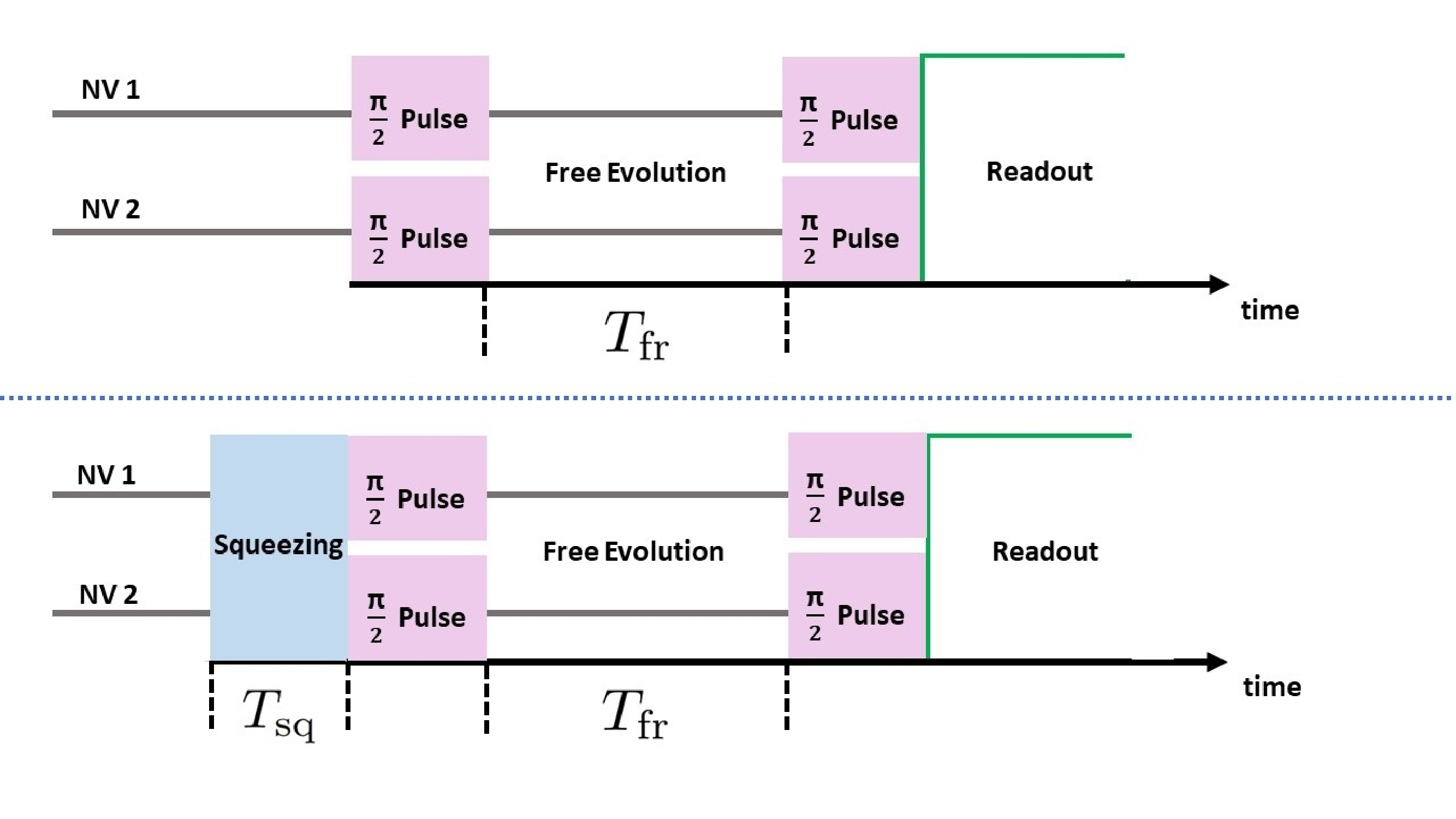}
		 \caption{The application of Ramsey-type sequence on two Nitrogen-Vacancy (NV) Centers with and without the introduction of spin squeezing. $T_{\mathrm{fr}}$ is the free evolution time. $T_{\mathrm{sq}}$ is the duration of the application of spin squeezing.}
        \label{Fig1}
\end{figure}

Initially, our focus is on a single NV center. We then extend our consideration to two NV centers, exploring Ramsey interferometry with and without the application of spin squeezing as illustrated in Fig.~\ref{Fig1}.

\subsection{Free Hamiltonian and Initial states}

Neglecting interactions with nuclear spins or spin-strain effects \cite{sewani2020coherent,ullah2022steady}, the Hamiltonian governing $N$ NV centers is expressed as follows,
\begin{equation}
    \hat H_0 = \sum_{i=1}^N \left [ D \hat S_{zi}^2 + g_S \vec{B} \cdot \vec{\hat S}_i \right ].
\end{equation}
We take $\hbar = 1$ here and in the rest of the paper. $D=2.87$~GHz is the temperature-dependent zero-field splitting, and $g_S=2.80$~MHz/G is the gyromagnetic ratio for the electron spin. The magnetic field taken parallel to the NV center axis ($z$) is denoted as $\vec{B} = (0, 0, B_z)$. We assume the NV axes are aligned with the magnetic field direction for simplicity, as is typical in NV center-based quantum sensing experiments using chemical vapor deposition methods \cite{osterkamp2019engineering,michl2014perfect}.
The spin-1 operator $\vec{\hat S}=(\hat S_x,\hat S_y,\hat S_z)$ is defined as
\begin{eqnarray}
\hat S_x &=& \frac{\hbar}{\sqrt{2}} \left( \begin{array}{ccc}
0 & 1 & 0 \\
1 & 0 & 1 \\
0 & 1 & 0 \end{array} \right), \\
\hat S_y &=& \frac{\hbar}{\sqrt{2}} \left( \begin{array}{ccc}
0 & -i & 0 \\
i & 0 & -i \\
0 & i & 0 \end{array} \right), \\
\hat S_z &=& \hbar \left( \begin{array}{ccc}
1 & 0 & 0 \\
0 & 0 & 0 \\
0 & 0 & -1 \end{array} \right).
\end{eqnarray}

$\hat S_{i}$ is the spin matrix for the $i^{th}$ NV center.
Due to the Zeeman effect in the presence of a magnetic field, the electron spin ($S=1$) is split into three states: $m_s=0$, $1$, $-1$, denoted as $\ket{0}$, $\ket{1}$ and $\ket{-1}$.
The initial state of the NV center is taken in the excited state $\ket{-1}$,
\begin{equation}
    \label{psi0}
    \ket{\psi_0} = \bigotimes_{i=1}^{N} \ket{-1}.
\end{equation}
This initialization can be achieved by first optically polarizing the spin into the $\ket{0}$ state, 
followed by a microwave $\pi$-pulse to transfer the population to $\ket{-1}$. 
Although this is a less common choice compared to starting in $\ket{0}$, 
initializing in $\ket{-1}$ is experimentally feasible \cite{macquarrie2015coherent} and physically motivated in our case, 
as the $\ket{-1}$ state is aligned along the $S_z$ axis, which is relevant for the dynamics considered in this work.
Then, each NV center undergoes a $\pi/2$ pulse, achieved by applying a rotation operator,
$$\mathrm{e}^{-i\frac{\pi}{2}\hat S_x}.$$
This pulse is experimentally generated by microwave drives \cite{mariani2020system}.

When incorporating spin squeezing, the initial state, $\ket{\psi_0}$, is first driven through spin squeezing, and then subjected to the $\pi/2$ pulse, as illustrated in Fig.~\ref{Fig1}.

\subsection{Spin Squeezing Generation}

To describe the intricate dynamics and capture the squeezing effects in the Hamiltonian for two NV centers, a squeezing term is introduced,
\begin{equation}
\label{sq1}
   \hat H_{\mathrm{S}}=  c_1 \left(\sum_{i=1}^N \hat S_{xi}\right)^2,
\end{equation}
where $c_1$ is the squeezing constant.
The Hamiltonian, as depicted in Eq.~(\ref{sq1}), is referred as a one-axis twisting Hamiltonian \cite{begzjav2021squeezing,wang2003spin}. The choice of $x$-axis twisting in the context of spin squeezing is strategic, as it allows for the manipulation of the noise cone, effectively narrowing the error ellipse of the spin. This, combined with the optimal initial state, enhances the precision of the magnetic field measurement by interplaying with the $z$-axis rotation induced by the magnetic field \cite{toth2009spin}.

The collective spin squeezing Hamiltonian in Eq.~(\ref{sq1}) explicitly includes interspecies interaction terms (e.g., $\hat{S}_{x1}\hat{S}_{x2}$ for $N=2$ NV centers), which are essential for generating collective spin squeezing. The following subsections discuss the methods to engineer these inter-NV interactions.

\subsubsection{Pulse Sequence Method}
A more direct approach to generating spin squeezing is through pulse sequences that induce an axis-twisting effect. In \cite{huang2021dynamic} a method for achieving spin squeezing via such pulse sequences is described in detail. Recent theoretical proposals and experimental efforts have further explored the realization of such Hamiltonians and the generation of spin squeezing using periodic pulse sequences in various quantum systems \cite{liu2023generation}. While their work focuses on spin-1/2 systems, this approach can be extended to spin-1 systems. This is particularly relevant to our setup, where we are interested in generating x-axis twisting. By adjusting the pulse properties, the method should allow us to transition from z-axis to x-axis twisting, which is more pertinent to the spin dynamics we are studying. A recalculation of the squeezing parameters within our parameter ranges will be necessary, but the method is sufficiently flexible to be implemented within our protocol.

\subsubsection{Electric Field Induced Strain Method}

An alternative method for generating spin squeezing, besides using pulse sequences, is the application of a strong electric field to induce strain in the system, leading to a Hamiltonian of the form $\sim [(S_x)^2 - (S_y)^2]$. Although this Hamiltonian differs from the one used in our work, it still develops spin squeezing along specific axes, as demonstrated in \cite{matsuzaki2022generation}.

\subsubsection{Off-resonant Field Mediated Effective Interaction Method}

Another approach, commonly used in spin-1/2 systems, involves effective interactions of the form $\sim S^+ S^- + h.c.$, mediated by an off-resonant field that is adiabatically eliminated, as outlined in \cite{macri2020spin}. To generalize this to spin-1 systems, one can introduce strong laser fields and replace photon operators with c-numbers, yielding a Hamiltonian of the form $\sim g(S^+)^2 + h.c.$ Alternatively, large detuning can be employed via the Schrieffer-Wolff (SW) transformation, also known as the Fröhlich transformation, as shown in \cite{pradana2021entanglement}, to eliminate photon degrees of freedom, leading to a simpler effective Hamiltonian. This method has also been applied in systems like silicon-vacancy centers, which are similar to NV centers, as described in \cite{ma2021preparation}.

\subsubsection{Other Methods}

The generation of this type of Hamiltonian in real setup has been further discussed in the literature. It can be realized through phonon mediation, as evidenced in \cite{PhysRevLett.110.156402} or produced in optical cavities \cite{borregaard2017one}. In our specific protocol, we need to deactivate spin squeezing after a specific duration. A dynamic control method for NV center spin squeezing, as discussed in \cite{song2017dissipation}, could potentially address the technical challenge.

Recent work \cite{block2024scalable} demonstrates the generation of large-scale spin squeezing from finite-temperature easy-plane magnetism, a method that shows great promise for NV centers.

To approximate a realistic measurement scenario, we calculate the Classical Fisher Information in~\ref{BB}.

\begin{figure}[t]
\centering

\setlength{\unitlength}{1cm}
\begin{picture}(0,0)
\put(-1.1,2.3){\makebox(0,0)[lt]{\textbf{(a)}}}
\end{picture}
\includegraphics[width=0.32\textwidth]{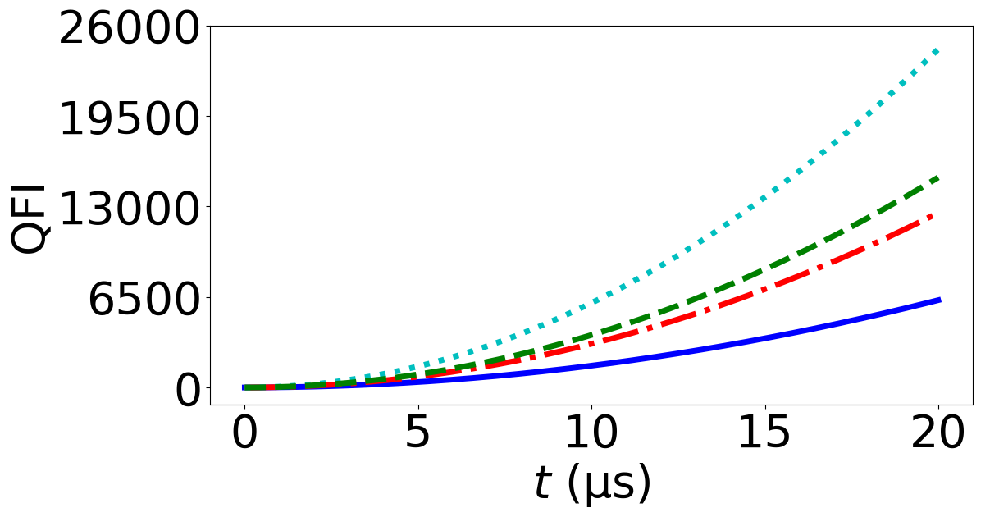}\hspace{1mm}

\begin{picture}(0,0)
\put(-1.1,2.3){\makebox(0,0)[lt]{\textbf{(b)}}}
\end{picture}
\includegraphics[width=0.32\textwidth]{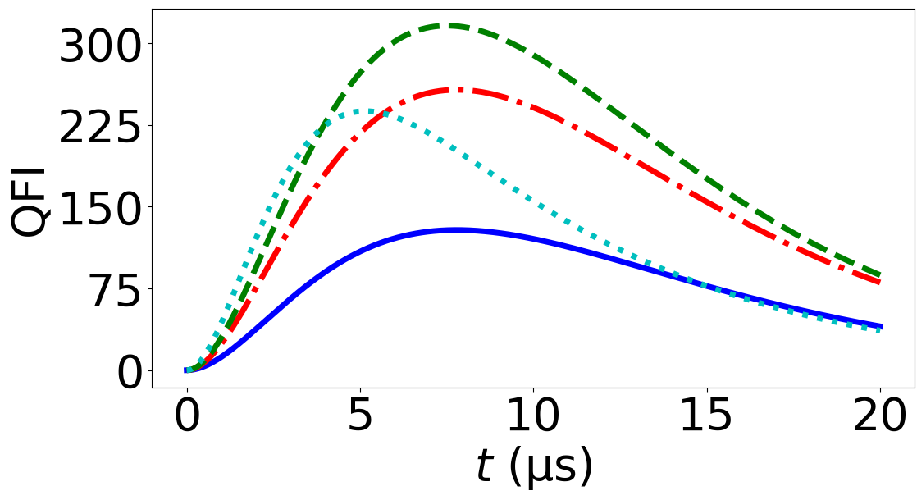}\hspace{1mm}

\begin{picture}(0,0)
\put(-1.1,2.3){\makebox(0,0)[lt]{\textbf{(c)}}}
\end{picture}
\includegraphics[width=0.32\textwidth]{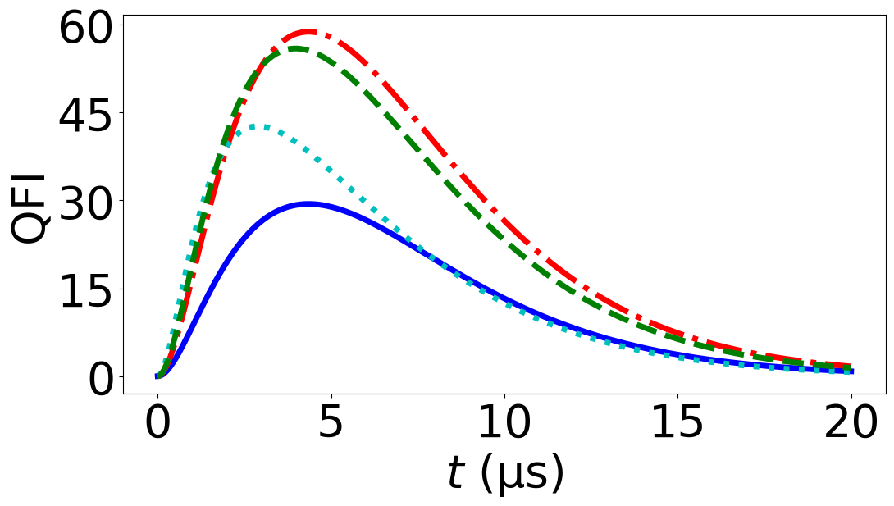}
\includegraphics[width=0.4\textwidth]{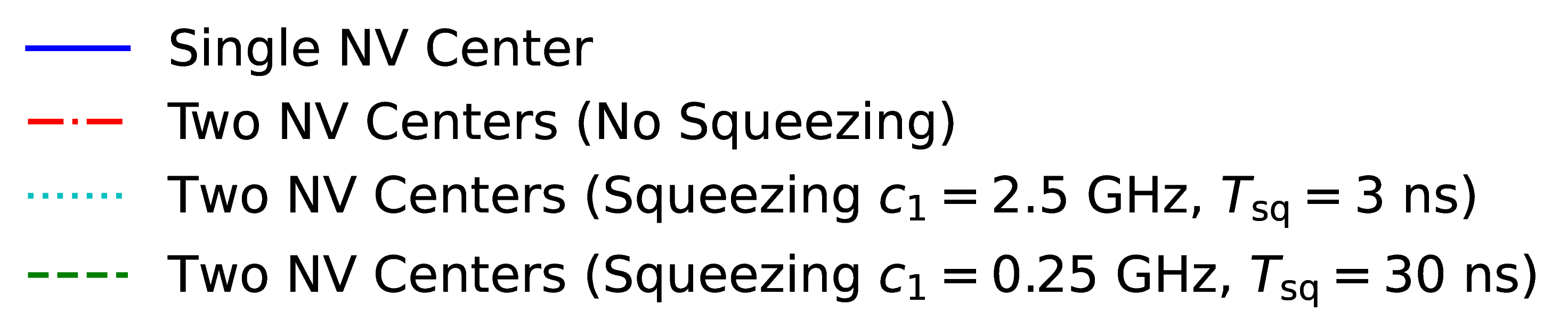}
\caption{QFI as a function of time with \( B = 50 \) G for the spin-1 formalism. The different scenarios are illustrated as: (a) without including any dissipation terms, (b) including thermalization and dephasing with \( \gamma_\mathrm{t} = 0.2 \) MHz and \( \gamma_\mathrm{d} = 0.1\gamma_\mathrm{t} \), and (c) including thermalization and dephasing with \( \gamma_\mathrm{t} = 0.2 \) MHz and \( \gamma_\mathrm{d} = \gamma_\mathrm{t} \). Curve identification is provided in the accompanying legend figure.}
\label{Fig2}
\end{figure}

\subsection{Dynamics}

We describe the open quantum system dynamics of the NV center by the following Lindblad master equation~\cite{breuer2002theory},
\begin{equation} \label{Eq::MasterEq}
\dot{\rho}(t) = -i[\hat H,\rho(t)] + \mathcal{D}_\mathrm{t}(\rho(t)) + \mathcal{D}_\mathrm{d}(\rho(t)),
\end{equation}
where the unitary contribution to the dynamics, $\hat H$, is provided as follows,
\begin{itemize}
    \item during free evolution by $\hat H=\hat H_0$,
    \item during spin squeezing by $\hat H=\hat H_0+\hat H_\mathrm{S}$.
\end{itemize}
The dynamics of the NV center system, including the effects of dephasing and relaxation, are modeled and numerically simulated using the Lindblad quantum master equation (Eq.~(\ref{Eq::MasterEq})). This approach allows for a comprehensive investigation of the system's behavior under various environmental conditions. All results presented in this work, including the comparative analysis of metrological capabilities, QFI calculations, and spin squeezing parameter evaluations, are derived from these numerical simulations. Specifically, the QFI is computed utilizing the \texttt{QuanEstimation} toolbox in Python, allowing for a systematic exploration of various parameters and environmental conditions, as detailed in the following subsections.

To model the system, we adopt a Markovian dissipation framework and analyze the interplay between dephasing and thermalization coefficients in optimizing the squeezing effect. Our approach employs a phenomenological master equation incorporating decoherence and dephasing terms, consistent with standard treatments of bulk NV centers, which rely on jump operators linear in spin-1 defect operators (\( S_+ \), \( S_- \), and \( S_z \)) \cite{PhysRevB.108.174418,PhysRevB.78.064309,PhysRevB.108.085203}. The NV center, while offering long coherence times and controllable spin dynamics, is nevertheless affected by the fluctuating states of nearby nuclear spins. These nuclear spins introduce a dynamic and unpredictable backdrop, contributing to the environmental dissipation that impacts the precision of measurements performed on the NV center. We assume very low temperatures relative to the NV center transition frequencies. Dephasing and thermalization are incorporated through the dissipators given as follows,
\begin{eqnarray}
\mathcal{D}_\mathrm{d}(\rho) &=& \sum_{i=1}^N \gamma_{\mathrm{d}} \left[\hat S_{zi}\rho(t)\hat S_{zi} - \rho(t)\right], \label{Eq::Dephase} \\
\mathcal{D}_\mathrm{t}(\rho) &=& \sum_{i=1}^N \gamma_\mathrm{t} \left[\hat S_i^- \rho(t) \hat S_i^+ - \frac{1}{2} \left\{ \hat S_i^+ \hat S_i^-, \rho(t) \right\} \right]. \label{Eq::Dissip}
\end{eqnarray}
where $\hat S_i^{-}$ is the lowering operator of the $i^{\mathrm{th}}$ NV center, and, 
$\hat S^{-}=\frac{1}{2}(\hat S_x-i \hat S_y)$. 
We define the thermalization rate, denoted by $\gamma_\mathrm{t}$, as 20~$\mu$s, based on the literature \cite{abeywardana2014magnetic}. For the dephasing rate, $\gamma_\mathrm{d}$, we will consider two scenarios: one where $\gamma_\mathrm{d}$ is 10$\%$ of the thermalization rate ($\gamma_\mathrm{d} = 0.1 \gamma_\mathrm{t}$), and another where it is equal to the thermalization rate ($\gamma_\mathrm{d} = \gamma_\mathrm{t}$).

In this study, we focus on NV centers located deep within the bulk of the diamond, at depths exceeding tens of nanometers from the crystal surface. This assumption minimizes the influence of surface effects, such as charge dissipation caused by near-surface defects, allowing us to concentrate on bulk NV center dynamics. While charge dissipation is generally more pronounced for near-surface NV centers, its contribution decreases significantly with increasing depth, as demonstrated in~\cite{candido2024interplay}. This reference provides a detailed quantitative analysis of the depth-dependent impact of charge and spin dissipation on decoherence and relaxation rates. For instance, it shows that the electric field fluctuations due to surface charges scale inversely with the square of the defect depth. At depths greater than tens of nanometers, the dissipation contribution from surface charges becomes negligible compared to bulk dissipation sources. These findings support our assumption that the charge dissipation for deep NV centers is substantially less impactful. Additionally, experimental results in~\cite{sangtawesin2019origins} confirm the depth dependence of relaxation and dephasing times, illustrating a marked improvement in coherence times as NV center depth increases. Figures 1 and 2 of~\cite{sangtawesin2019origins} highlight this trend, with T2 and T1 coherence times improving by approximately two orders of magnitude as the depth increases from 4 nm to 12 nm.

\section{Quantum Measures}
\label{fisher}
\subsection{Quantum Fisher Information}

In the context of quantum parameter estimation, particularly for magnetic field sensing, fundamental results in quantum metrology become imperative. Let us consider a scenario where a quantum state encodes a parameter $\theta$, and the goal is to estimate this parameter with optimal precision~\cite{luo2000quantum,liu2020quantum}.
For a mixed state described by the density operator, $\hat{\rho}$, the QFI can be defined, using the Symmetric Logarithmic Derivative (SLD) operator ($\hat{L}_\theta$), as
\begin{equation}
F_Q(\theta) = \mathrm{Tr}[\hat{\rho} \hat{L}_\theta^2].
\end{equation}
The SLD operator is implicitly determined by the equation,
\begin{equation}
\partial_\theta \hat{\rho} = \frac{\hat{L}_\theta \hat{\rho} + \hat{\rho} \hat{L}_\theta}{2}.
\end{equation}
Expressing $\hat{L}_\theta$ in the eigenbasis of $\hat{\rho}$, the QFI takes the form,
\begin{equation}
F_Q(\theta) = 2 \sum_{k, l} \frac{\left| \langle k | \partial_\theta \hat{\rho} | l \rangle \right|^2}{\lambda_k + \lambda_l}.
\end{equation}
Here, $\ket{k}$ and $\ket{l}$ are the eigenvectors and $\lambda_k$ and $\lambda_l$ are the eigenvalues of the density operator $\hat{\rho}$.

The Quantum Fisher Information (QFI) sets a lower bound on the uncertainty of parameter estimation through the Cramér-Rao bound:
\[
\Delta \theta \geq \frac{1}{\sqrt{F_Q(\theta)}}.
\]
This expression shows that increasing the QFI improves the precision in estimating the parameter \(\theta\).
Even in the absence of spin squeezing, one can motivate the use of higher-dimensional quantum systems through a simple argument. Consider a Zeeman interaction of the form \( H = -\gamma B S_z \). For a spin-1/2 system, the QFI is \( F_Q = 4\gamma^2 \), where the factor of 4 can be associated with the structure of the two-dimensional Hilbert space. Extending this to a spin-1 (qutrit) system, one intuitively expects \( F_Q = 9\gamma^2 \), reflecting the larger separation between eigenvalues of \( S_z \) and the increased dimensionality of the system. In such cases, the optimal measurement is along the spin direction aligned with the magnetic field—typically the \( z \)-axis component. While simply using a qutrit provides some advantage, spin squeezing offers a strategy to harness the full potential of the higher-dimensional Hilbert space, analogous to how GHZ states or optical squeezed states are used to approach the Heisenberg limit in other platforms.
Thus, although the expanded Hilbert space of spin-1 systems opens up greater possibilities for enhanced sensitivity, reaching the Heisenberg limit generally requires careful state preparation and the use of entanglement or squeezing techniques.

In the specific application of estimating a magnetic field, $\theta$, in our case, is the strength of the magnetic field, and the QFI provides a quantitative measure of the ultimate precision achievable in such quantum magnetic field sensing experiments. This theoretical foundation is critical for understanding the limits and potential advancements in quantum-enhanced magnetic field measurements.
To quantify this precision, we employ the \texttt{QuanEstimation} toolbox in Python \cite{zhang2022quanestimation} to calculate the QFI.

\subsection{Spin Squeezing Parameter}
\label{Sq}

The concept of spin squeezing is established by defining it as the condition where the variance of a spin component normal to the mean spin is smaller than the standard quantum limit. To quantify the extent of spin squeezing, we employ the spin squeezing parameter, as introduced by Kitagawa and Ueda \cite{kitagawa1993squeezed}. Kitagawa-Ueda spin squeezing brings quantum entanglement perspective to the usual spin squeezing which is the reduction of spin noise in a certain direction at the cost of more noise in other directions. This criterion is given by
\begin{equation}
    \xi^2 = \frac{4(\Delta \hat{J}^2_{n_{\perp}})_{\mathrm{min}}}{N},
\end{equation}
where $n_{\perp}$ is an axis perpendicular to the mean-spin direction $\hat n$, $N$ is the number of spins (in our case, $N=2$), and $\hat{J}$ represents the collective spin operator.

In this work, we are using the particle-entanglement-based definition of spin squeezing, which is more restrictive than the noise redistribution version introduced by Wineland \cite{wineland1994squeezed}, commonly used in Ramsey measurements. The entanglement-based definition requires true multi-particle entanglement and cannot be achieved with a single spin, whereas the Wineland definition can reflect spin squeezing even in the absence of particle entanglement, merely by redistributing noise between different spin components.

Kitagawa-Ueda and other quantum correlations based criteria of spin squeezing \cite{ma2011quantum} are commonly defined for SU(2) or spin-1/2 atomic ensembles. Spin squeezing in SU(3) or spin-1 systems can be discussed in their SU(2) subgroups, such as isospin squeezing \cite{mustecapliouglu2002spin}. We follow a similar approach to characterize quantum correlations and spin squeezing within the manifold of $\{\ket{0},\ket{-1}\}$. In general, spin squeezing can be optimized by rotating the coordinate frame and choosing the critical rotation angles. For simplicity, we only consider a fixed reference frame. Based on this manifold, the spin matrices $\hat \sigma=(\hat \sigma_x,\hat \sigma_y,\hat \sigma_z)$ are written as
\begin{eqnarray}
\hat \sigma_x &=& \ket{0}\bra{-1} + \ket{-1}\bra{0}, \\
\hat \sigma_y &=& i\left( \ket{0}\bra{-1} - \ket{-1}\bra{0} \right), \\
\hat \sigma_z &=& \ket{-1}\bra{-1} - \ket{0}\bra{0}.
\end{eqnarray}
and the collective spin $\hat J=(\hat J_x,\hat J_y,\hat J_z)$ is defined as
\begin{eqnarray}
    \hat J_x&= \frac{1}{2}(\hat\sigma_x\otimes I_3 + I_3\otimes \hat\sigma_x),\\
    \hat J_y&= \frac{1}{2}(\hat\sigma_y\otimes I_3 + I_3\otimes \hat\sigma_y),\\
    \hat J_z&= \frac{1}{2}(\hat\sigma_z\otimes I_3 + I_3\otimes \hat\sigma_z),
\end{eqnarray}
with $I_3$ being the three-dimensional unit matrix. To calculate spin squeezing, we calculate the mean spin squeezing direction $\hat n$ for the state,
\begin{equation}
    \hat n=\frac{( \langle \hat J_x \rangle,\langle \hat J_y \rangle,\langle \hat J_z \rangle)}{|\langle \hat J \rangle|}.
\end{equation}
And then, we take the minimal variance, $(\Delta \hat{J}^2_{n_{\perp}})_{\mathrm{min}}$, in the perpendicular direction of $\hat n$. The presence of spin squeezing is characterized by $\xi^2<1$.
Spin squeezing is closely tied to the creation of entanglement~\cite{toth2009spin,sirsi2004spin,vitagliano2014spin}. Additionally, a previous study suggests that initial squeezing can aid in entanglement~\cite{pradana2021entanglement}. Our results are consistent with this statement (\ref{AA}).

Our approach highlights the quantum metrological advantage of employing higher-dimensional (spin-1) multipartite quantum entanglement. This advantage, distinct from conventional spin noise squeezing, is achieved by leveraging entanglement-based criteria for spin squeezing to further enhance sensitivity in quantum measurement protocols.

While we evaluate spin squeezing within the $\{|0\rangle, |-1\rangle\}$ subspace for metrological relevance, the full spin-1 (qutrit) Hilbert space is retained in our model. The state $\ket{+1}$ and its couplings remain active in the dynamics and decoherence, distinguishing the qutrit from the reduced qubit model and accounting for their differing behavior.

\begin{figure}[t!]
		\centering
   		\includegraphics[width=0.9\linewidth]{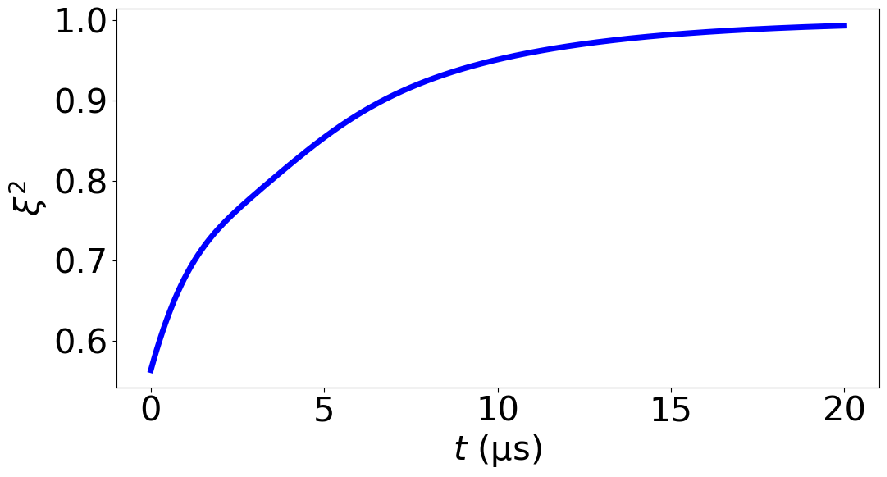}
		\caption{For the spin-1 formalism, the curve in solid blue represents the squeezing parameter, $\xi^2$, as a function of time during the free evolution of two NV centers subjected to spin squeezing with $c_1 = 2.5$~GHz for a time of $T_{\mathrm{sq}} = 3$~ns, including thermalization and dephasing with $\gamma_\mathrm{t} = 0.2$~MHz, $\gamma_\mathrm{d} = 0.1\gamma_\mathrm{t}$, and $B = 50$~G.}
        \label{Fig3}
\end{figure}

\section{Results}
\label{results}
In our simulations, the magnetic field was fixed at $B=50$~G, but similar results were obtained across a range of values from $5$ to $500$~G.
Initially, we investigate the typical Ramsey sequence for one and two NV centers, both starting in the initial state (Eq.~\ref{psi0}). Subsequently, after a $\pi/2$ pulse, we allow the system to undergo evolution through the free Hamiltonian $\hat H_0$ for a duration of $T_{\mathrm{fr}} = 20$ $\mu$s,  $T_{\mathrm{fr}}$ being the free evolution time. We choose this time taking into account the dissipative time that falls within the order of tenth of microseconds.
To further enhance precision, we start again in the initial state (Eq.~\ref{psi0}), however before the $\pi/2$ pulse and the free evolution, we introduce spin squeezing for a duration of $T_{\mathrm{sq}}$ which corresponds to the time it takes for the squeezing parameter to reach its minimum value. The two scenarios are illustrated in Fig.~\ref{Fig1}.

\subsection{Analysis with spin-1 formalism}

Fig.~\ref{Fig2} compares one NV center without spin squeezing (blue), two NV centers without spin squeezing (red) and two NV centers with spin squeezing applied at varying strengths, specifically with squeezing constants of $c_1=0.25$ GHz (green) and $c_1=2.5$ GHz (cyan).

In the case without spin squeezing, Fig.~\ref{Fig2}(a) shows a twofold increase in the QFI when transitioning from one NV center to two NV centers. When environmental dissipation is introduced (Fig.~\ref{Fig2}(b) and Fig.~\ref{Fig2}(c)), whether the QFI increases or decreases for one NV center, the twofold enhancement persists for two NV centers.

In the case with spin squeezing, in the absence of dissipation (Fig.~\ref{Fig2}(a)), the QFI is larger when spin squeezing is applied with squeezing constant, $c_1=0.25$ GHz. It increases further with a higher squeezing constant, $c_1=2.5$ GHz.
When environmental dissipation is introduced with \( \gamma_{\mathrm{d}} = 0.1\gamma_{\mathrm{t}} \) (Fig.~\ref{Fig2}(b)), spin squeezing remains beneficial, particularly at a lower squeezing constant, $c_1=0.25$ GHz. This suggests that in the presence of dissipation, a higher squeezing constant does not necessarily lead to higher QFI. Instead, a low squeezing constant ($c_1=0.25$ GHz), which is sufficient enough to generate spin squeezing, is more advantageous. However, within the first 4 \(\mu\)s, a higher squeezing constant, $c_1=2.5$ GHz, provides more benefit.

Finally, Fig.~\ref{Fig2}(c) shows that the advantage of squeezing is lost when the dephasing rate increases to \( \gamma_{\mathrm{d}} = \gamma_{\mathrm{t}} \). While spin squeezing enhance magnetic sensing capabilities, they are also highly sensitive to environmental dissipation, particularly dephasing. Analyzing the behavior of the QFI curves under such dissipation conditions is crucial. By identifying the time at which the QFI reaches its maximum, we can optimize experimental design and ensure measurements are performed at the peak sensitivity. Experimentally, we typically measure populations or ($\hat S_z$) values. However, it is important to note that while the QFI is calculated for the optimal Positive Operator-Valued Measure (POVM), specifically the eigenprojectors of the SLD, measuring $\hat S_z$ could potentially be suboptimal. In practical NV center experiments, measurements are typically performed by projecting the system into the \(S_z\) basis. This is achieved by applying a second \(\pi/2\)-pulse at the end of the free evolution time, which rotates the accumulated phase from the transverse plane (\(S_x, S_y\)) onto the \(S_z\)-axis, where population differences are then measured.
Our method is designed for measuring DC magnetic fields, where the typical readout technique is optically detected magnetic resonance (ODMR), which relies on monitoring changes in photoluminescence intensity under resonant microwave excitation \cite{rondin2014magnetometry, oon2022ramsey, alkauskas2014first, matsuzaki2016optically}.

The effect of spin squeezing on the NV centers is further illustrated in Fig.~\ref{Fig3}. The squeezing parameter, initially around 0.5 (below 1) at the beginning of free evolution, increases due to dephasing and thermalization during this period.~\ref{AA} demonstrates that spin squeezing also leads to entanglement.

\begin{figure}[t!]
		\centering
   		\includegraphics[width=0.9\linewidth]{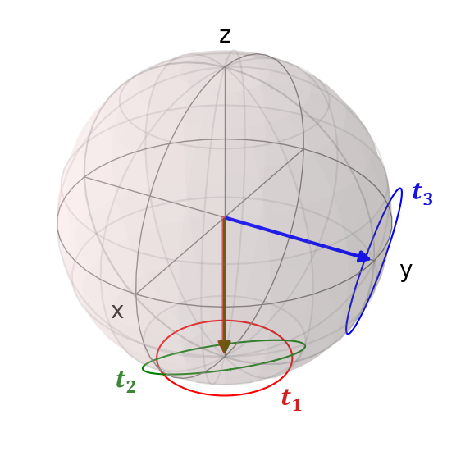}
		\caption{Illustration of the mean spin orientation and the variance of state preparation before the free evolution of the NV centers. At time \( t_1 \), the state is in its initial configuration (Eq.~(\ref{psi0})) with homogeneous variance. At time \( t_2 \), the state reaches the end of the spin squeezing period, remaining aligned in the \( z \)-direction but with noise squeezed into an ellipse. Finally, at \( t_3 \), a \( \pi/2 \) pulse is applied, shifting the state into the \( x \)-\( y \) plane.}
        \label{Fig4}
\end{figure}

The spin system is initially oriented along the \( z \)-axis. The expectation values of the angular momentum operators are
\begin{equation}
\langle J_z \rangle = \frac{N}{2}, \quad \langle J_x \rangle = 0, \quad \langle J_y \rangle = 0,
\end{equation}
where \( N \) is the total number of spins in the ensemble.
In this state, the spin variance is symmetric around the \( z \)-axis, forming a circular uncertainty distribution in the \( x \)-\( y \)-plane. The uncertainty in each component is
\begin{equation}
\langle (\Delta J_x)^2 \rangle = \langle (\Delta J_y)^2 \rangle = \frac{N}{4}.
\end{equation}
When spin squeezing is applied, the variance along one axis in the \( x \)-\( y \) plane is reduced, while the variance along the orthogonal axis increases. The squeezed variances become,
\begin{equation}
\langle (\Delta J_\mathrm{sq})^2 \rangle = \frac{N}{4\xi^2}, \quad \langle (\Delta J_\mathrm{anti-sq})^2 \rangle = \frac{N\xi^2}{4},
\end{equation}
where \( \xi^2 < 1 \) is the squeezing parameter. When spin squeezing is applied, the spin variance is reduced along a specific direction in the \( x \)-\( y \) plane and increased along the orthogonal direction. Although we do not analytically specify the squeezed axis, it is uniquely determined for each state as the direction minimizing the spin variance perpendicular to the mean spin vector. This squeezed axis lies within the transverse plane due to our choice of squeezing Hamiltonian. After the \(\pi/2\) pulse, the squeezed axis is effectively aligned with the direction in which the spin state accumulates phase due to the magnetic field, ensuring that the reduced uncertainty translates into enhanced phase sensitivity.
After applying a \(\pi/2\) pulse, the spin state is rotated into the \( x \)-\( y \) plane, and the noise ellipse is now aligned in this plane. This sequence is illustrated in Fig.~\ref{Fig4}. The pulse does not alter the relative squeezing, but just the orientation of the state. The effect of the magnetic field \( B \), applied along the \( z \)-axis, is to precess the spin state around the \( z \)-axis.

While we did not adopt a non-exponential decoherence model, such an approach could provide valuable insights by capturing slower, more realistic dissipation dynamics \cite{choi2017depolarization}. Furthermore, non-exponential decay could potentially extend coherence times, thus enhancing metrological precision in practical applications. Investigating this further may increase the relevance of our results to real-world quantum systems.

Moreover, we examine non-Markovian effects by introducing an ancillary qubit into the model, resulting in improved metrological performance, as discussed in Section \ref{NonMar}.

\begin{figure}[t]
\centering
\setlength{\unitlength}{1cm}
\begin{picture}(0,0)
\put(-1.1,2.3){\makebox(0,0)[lt]{\textbf{(a)}}}
\end{picture}
\includegraphics[width=0.32\textwidth]{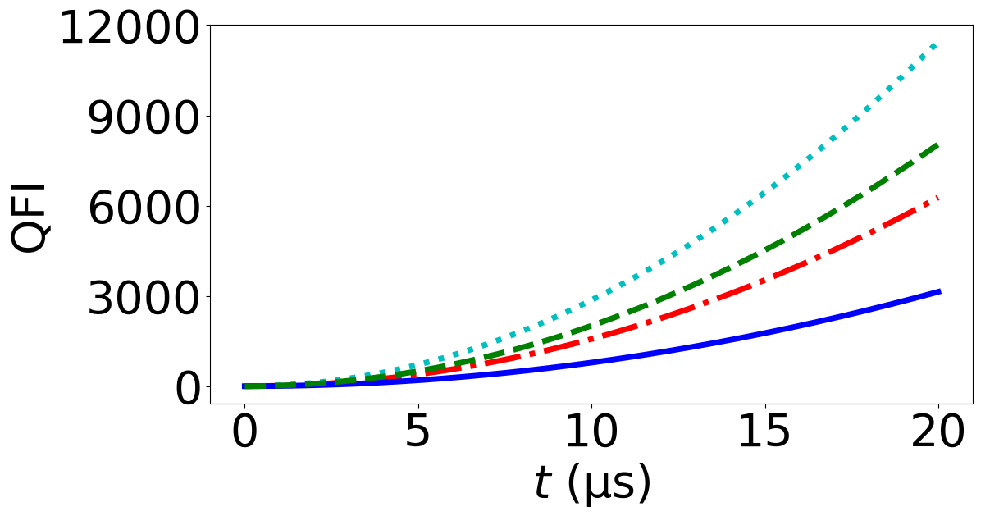}\hspace{1mm}

\begin{picture}(0,0)
\put(-1.1,2.3){\makebox(0,0)[lt]{\textbf{(b)}}}
\end{picture}
\includegraphics[width=0.32\textwidth]{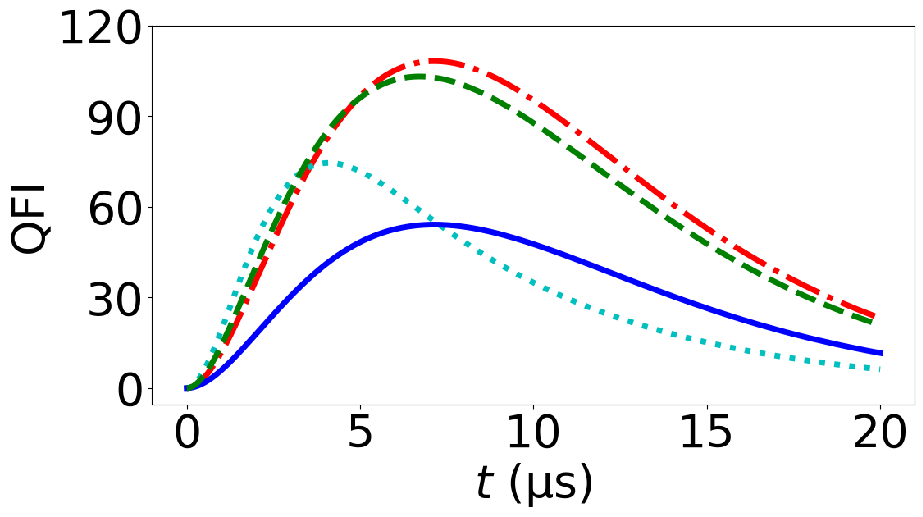}\hspace{1mm}

\begin{picture}(0,0)
\put(-1.1,2.3){\makebox(0,0)[lt]{\textbf{(c)}}}
\end{picture}
\includegraphics[width=0.32\textwidth]{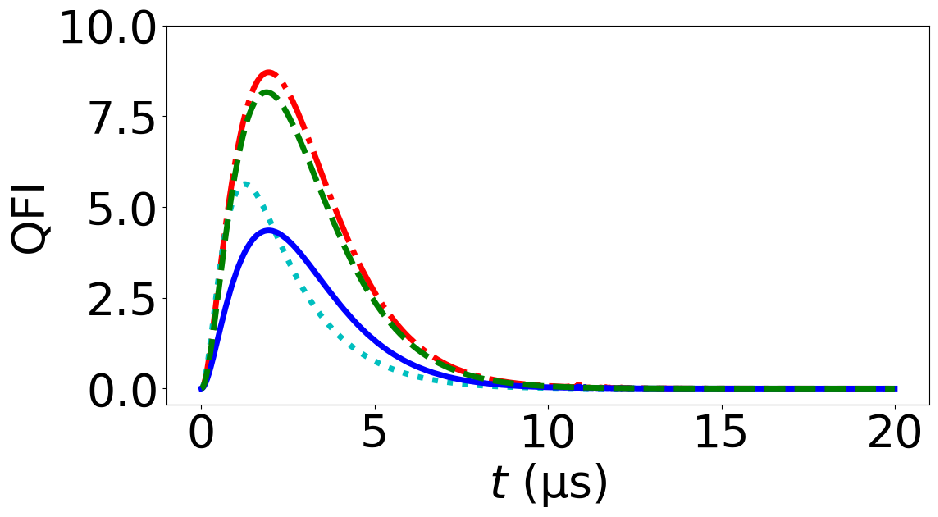}
\includegraphics[width=0.4\textwidth]{legend_2_5.eps}
\caption{QFI as a function of time with \( B = 50 \) G for the spin-1/2 formalism. The different scenarios are illustrated as: (a) without including any dissipation terms, (b) including thermalization and dephasing with \( \gamma_\mathrm{t} = 0.2 \) MHz and \( \gamma_\mathrm{d} = 0.1\gamma_\mathrm{t} \), and (c) including thermalization and dephasing with \( \gamma_\mathrm{t} = 0.2 \) MHz and \( \gamma_\mathrm{d} = \gamma_\mathrm{t} \). Curve identification is provided in the accompanying legend figure.}
\label{Fig5abc}
\end{figure}

\subsection{Analysis with spin-1/2 formalism}

In this section, we examine the outcomes within the identical parameter scope for the spin-1/2 framework. This is achieved by concentrating on the dynamics of the states $\ket{0}$ and $\ket{-1}$, which correspond to the standard qubit subspace commonly used in NV center-based Ramsey interferometry. This configuration effectively represents the regular NV Ramsey scheme, allowing for a direct and meaningful comparison with our spin-1 (qutrit) results under comparable conditions, including the same dissipation parameters. The Hamiltonian, in this context, is characterized as follows,
\begin{equation}
 \hat H_0 =  \sum_{i=1}^N \frac{\omega}{2} \hat \sigma_{zi},
\end{equation}
where $\omega=D-g_S B_z$ and $\hat \sigma_i=(\hat \sigma_{xi},\hat \sigma_{yi},\hat \sigma_{zi})$ are the Pauli matrices for the $i^{\mathrm{th}}$ NV center.
The parameters and open quantum system dynamics are kept same as in the spin-1 model. Eq.~(\ref{Eq::Dephase}) and Eq.~(\ref{Eq::Dissip}) are modified respectively to
\begin{eqnarray}
\mathcal{D}_\mathrm{d}(\rho) &= \sum^N_{i=1} \gamma_{\mathrm{d}} \,[\hat \sigma_{zi}\rho(t)\hat \sigma_{zi} -\rho(t)]; \label{Eq::Dephase2}\\
\mathcal{D}_\mathrm{t}(\rho) &=  \sum^N_{i=1} \gamma_{\mathrm{t}} \, [\hat \sigma_i^-\rho(t)\hat \sigma_i^{+} -\frac{1}{2} \left\{\hat \sigma_i^{+}\hat \sigma_i^-,\rho(t)\right\}]\label{Eq::Dissip2},
\end{eqnarray}
where $\hat \sigma_i^{\pm}$ are the Pauli spin ladder operators for the $i^{\mathrm{th}}$ NV center.
The spin squeezing Hamiltonian in Eq.~(\ref{sq1}) is modified to
 \begin{equation}
    \hat H_{\mathrm{S}}= c_1 \left(\sum^N_{i=1} \hat \sigma_{xi}\right)^2.
\end{equation}
The initial state of the NV center is taken in the excited state $\ket{-1}$. Then each NV center undergoes an $\pi/2$ pulse, achieved by applying a rotation operator,
$$\mathrm{e}^{-i\frac{\pi}{2}\hat \sigma_x}.$$
As in the spin-1 formalism, if spin squeezing is added, it is applied before the $\pi/2$ pulse.
We then proceed to analyze the QFI under both scenarios (with and without squeezing), considering the presence and absence of dissipation.

In Fig.~\ref{Fig5abc}(a), we observe that without considering any dissipation channel, spin squeezing yields a higher QFI. However, in the presence of dissipation [Fig.~\ref{Fig5abc}(b) and Fig.~\ref{Fig5abc}(c)], the advantages of spin squeezing are lost even for small dephasing [Fig.~\ref{Fig5abc}(b)], and Ramsey sequence without spin squeezing is more beneficial. Earlier in Fig.~\ref{Fig2}(b), for the qutrit case, we observed advantages even in the presence of dephasing. Therefore, the benefits of using qutrit sensors over qubit sensors are clearly evident.

\subsection{Non-Markovian Environment} \label{NonMar}

We simulate non-Markovian effects by introducing an ancillary qubit coupled to a single NV center, effectively forming a minimal finite environment~\cite{PhysRevA.95.012122, PhysRevLett.121.060401}. This coupling, implemented via an \( XX + YY \) interaction with a strength of 0.1 GHz, enables bidirectional information exchange between the system and the environment, a hallmark of non-Markovian dynamics~\cite{breuer2016colloquium}. Such memory effects manifest as deviations from purely exponential decay, including coherence revivals and enhanced QFI.

Both the NV center and the ancillary qubit are subject to environmental decoherence, including dephasing and thermalization. However, the NV center now experiences these effects indirectly through its interaction with the ancillary spin. The finite size of this spin environment allows information to flow back into the NV center, reproducing non-Markovian characteristics without requiring an explicit spectral density model. Moreover, the presence of the ancillary qubit modifies the effective bath spectral response seen by the NV center, rendering it Lorentzian with a finite bandwidth. This spectral structure leads to non-Markovian open system evolution, as discussed in~\cite{aiache2025quantum}.

\begin{figure}[t!]
\centering
\includegraphics[width=0.9\linewidth]{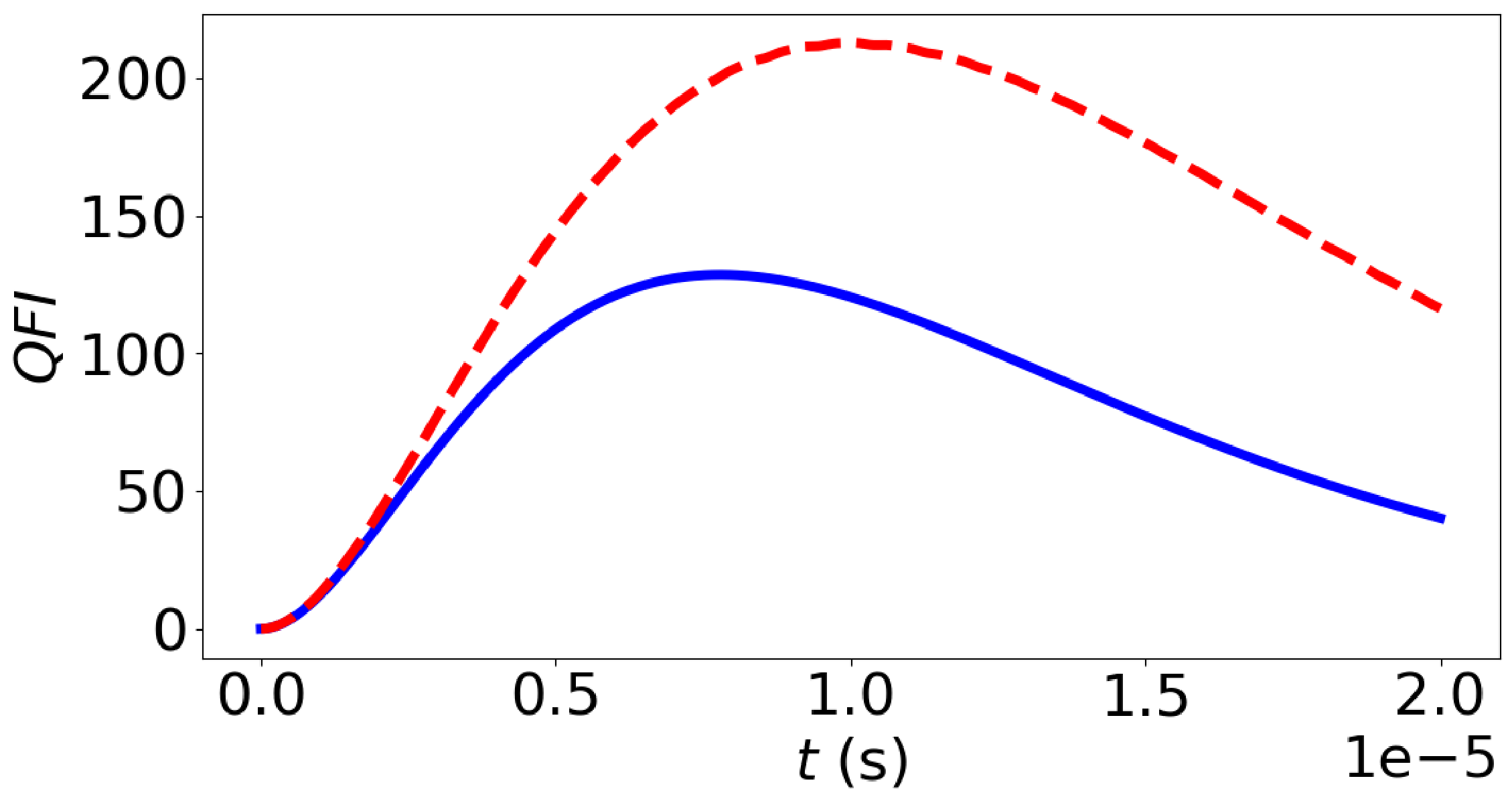}
\caption{For the spin-1 formalism, the curves show the quantum Fisher information (QFI) as a function of time during the free evolution of a single NV center, including thermalization and dephasing with $\gamma_\mathrm{t} = 0.2$ MHz, $\gamma_\mathrm{d} = 0.1\gamma_\mathrm{t}$, and $B = 50$ G. The blue solid curve represents QFI under Markovian dissipation, while the red dashed curve shows QFI under non-Markovian dissipation.
}
\label{Fig6}
\end{figure}

As shown in Fig.~\ref{Fig6}, this configuration leads to significantly higher QFI, demonstrating the metrological advantage of non-Markovian regimes. By introducing just one ancillary qubit, we create a structured environment that shifts the system's response from a Markovian to a non-Markovian regime.

In practice, NV centers are often surrounded by a finite number of nuclear or impurity spins, which naturally form non-Markovian environments. Our simplified model captures these effects and illustrates the potential of engineered non-Markovian dynamics to enhance quantum sensing~\cite{el2024non,haase2018controllable}.

\section{Conclusion}
\label{conclusion}
The study concludes that incorporating a spin squeezing step in magnetometric measurement and utilizing qutrit NV centers, as demonstrated through our comprehensive numerical simulations, yields a precision increase relative to standard Ramsey magnetometry. These findings are strongly supported by comprehensive simulations across diverse parameters and environmental settings. While the advantage of spin squeezing is notably sensitive to the dephasing environment, becoming marginal with increased dephasing, our simulations showed that moderate squeezing resulted in higher QFI compared to strong squeezing over extended periods. The analysis, computationally comparing spin-1/2 and spin-1 formalisms, unequivocally demonstrated the advantages of using a qutrit system over a qubit system.

In light of surface effects, the dominant role of charge dissipation near the surface severely challenges the exploitation of spin squeezing and multipartite entanglement for magnetometry. Addressing these limitations requires exploring alternative strategies to enable quantum entanglement advantages for near-surface probes. Future research could focus on identifying optimal depths and surface properties to enhance magnetometry precision under non-Markovian dynamics and spin squeezing. These efforts are crucial for unlocking the full potential of NV centers for advanced sensing applications. 

Additionally, we have observed that spin squeezing is typically associated with multipartite entanglement under certain conditions, but a detailed quantification of entanglement was not performed in this study and remains a subject for future work. A more detailed investigation could help clarify whether the observed enhancement in precision is primarily due to the redistribution of spin noise or the presence of entanglement.

Our findings contribute to the broader exploration of quantum technologies beyond qubits, where qutrits and higher-dimensional qubits (qudits) are increasingly considered, particularly in the context of quantum computing \cite{ogunkoya2024qutrit,xu2024experimental}. An intriguing question persists regarding the scalability, as outlook in \cite{adani2024critical}, of these advantages, especially regarding the utilization of NV center qutrit ensembles or the investigation of higher-level systems (qudits). These avenues pose exciting challenges for future studies, promising further insights.\\

\section*{Acknowledgements}

This work was supported by the Scientific and Technological Research Council of Türkiye (TÜBITAK) under Project Number 123F150. L.~G.~and Ö.~E.~M.~thank TÜBITAK for their support.

\appendix

\section{Entanglement Measure}
\label{AA}

\begin{figure}[t!]
		\centering
		\includegraphics[width=0.9\linewidth]{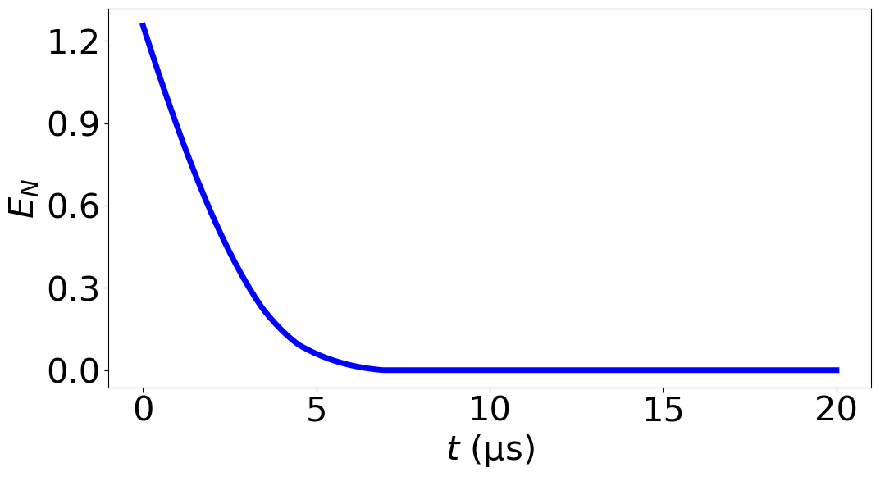}
		 \caption{For the spin-1 formalism, the curve in solid blue represents the logarithmic negativity, $E_N$, as a function of time during the free evolution of two NV centers subjected to spin squeezing with $c_1 = 2.5$ GHz for a time of $T_{\mathrm{sq}} = 3$ ns, including thermalization and dephasing with $\gamma_\mathrm{t} = 0.2$ MHz, $\gamma_\mathrm{d} = 0.1\gamma_\mathrm{t}$, and $B = 50$ G.}
		 \label{FigA1}
\end{figure}

For higher-dimensional systems than qubits, some entanglement measures exist. The negativity, denoted as $N(\rho_{AB})$, quantifies entanglement in bipartite quantum systems $\rho_{AB}$ \cite{benabdallah2021dynamics}. It is defined as the absolute sum of the negative eigenvalues of the partially transposed density matrix $\rho_A^{T}$,
\begin{equation}
N(\rho_{AB}) = \sum_i |\lambda_i|,
\end{equation}
where $\lambda_i$ are the negative eigenvalues of the density matrix $\rho_A^{T}$.
The logarithmic negativity, denoted as $E_N(\rho_{AB})$, is related to the negativity and serves as a good indicator of the degree of entanglement in our system. It is defined as
\begin{equation}
 E_N(\rho_{AB}) = \log_2(2N(\rho_{AB}) + 1).
\end{equation}

In Fig.~\ref{FigA1}, we observe that the application of spin squeezing to the two NV centers system creates an entanglement between them. This entanglement is illustrated by an increase in the logarithmic negativity during the application of spin squeezing. During free evolution, this entanglement decreases due to environmental dissipation. Thus, the creation of spin squeezing in our spin-1 systems is accompanied by the creation of entanglement. We emphasize that such multipartite entanglement arises naturally as a consequence of spin squeezing, where the NV centers are treated as effectively interacting through a collective spin-squeezing mechanism. Multipartite entanglement can be a byproduct of spin squeezing, and spin squeezing is the main resource we harness for metrological advantage here. We do not focus on the implementation of an entanglement protocol for spatially separated NV centers here, such as those based on dipolar interaction or photon-mediated schemes; instead, in the main text, we discuss how spin squeezing can possibly be generated between two NV centers.

\section{Classical Fisher Information}
\label{BB}

We calculate the Classical Fisher Information (CFI) during the system's free evolution to better reflect a realistic experimental setup. Specifically, we evaluate the CFI using measurements along the eigenvectors of $S_x$ and $S_y$, which can be practically implemented through state tomography to reconstruct the quantum state and analyze its sensitivity to parameter changes.
In quantum metrology, the CFI is a valuable metric for quantifying how responsive a measurement is to variations in a system parameter. For a probability distribution $\{ p_i(\theta) \}$ that depends on a parameter $\theta$, the CFI is given by

\begin{figure}[t!]
		\centering
		\includegraphics[width=0.9\linewidth]{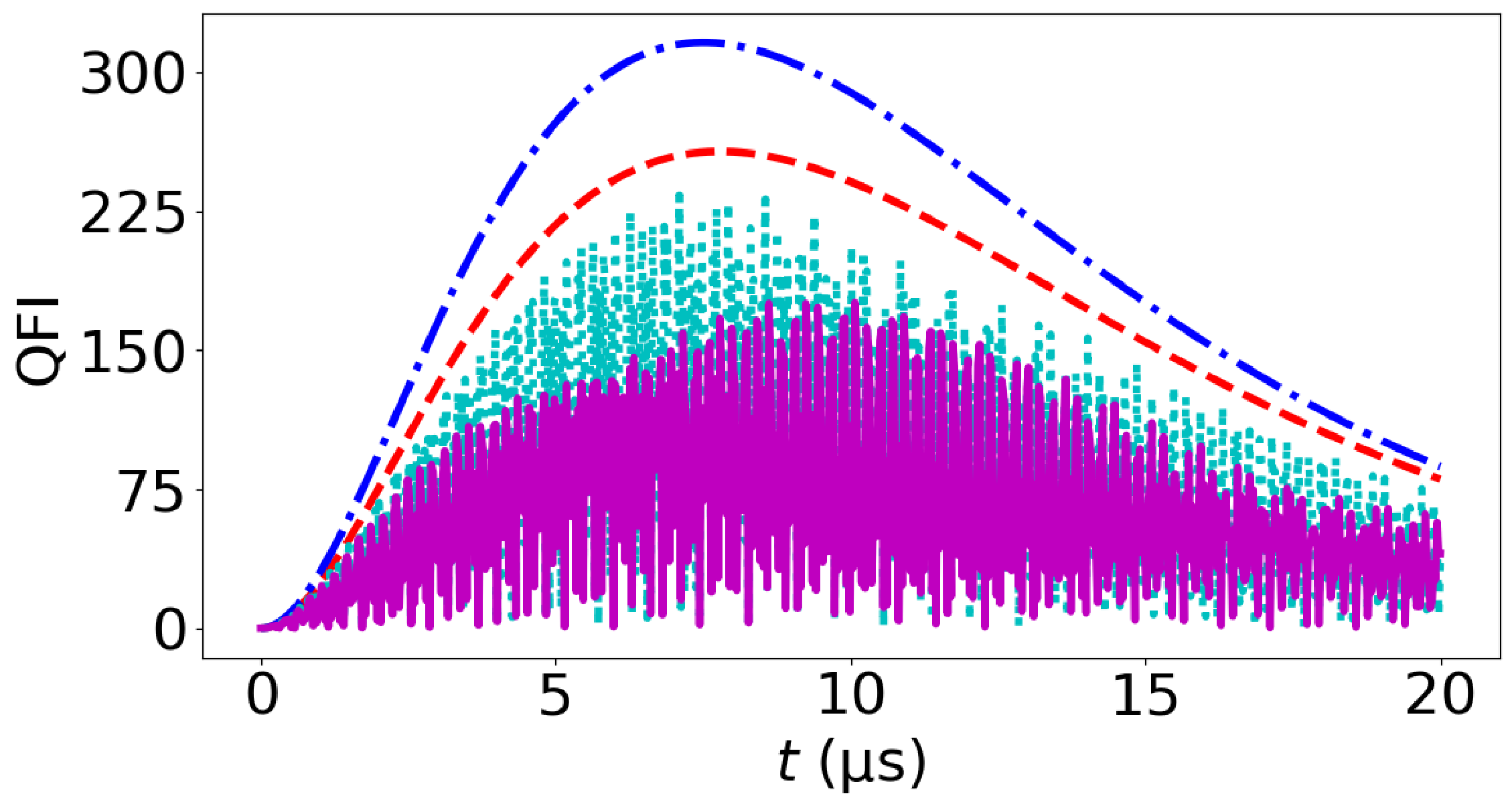}
        \includegraphics[width=0.4\textwidth]{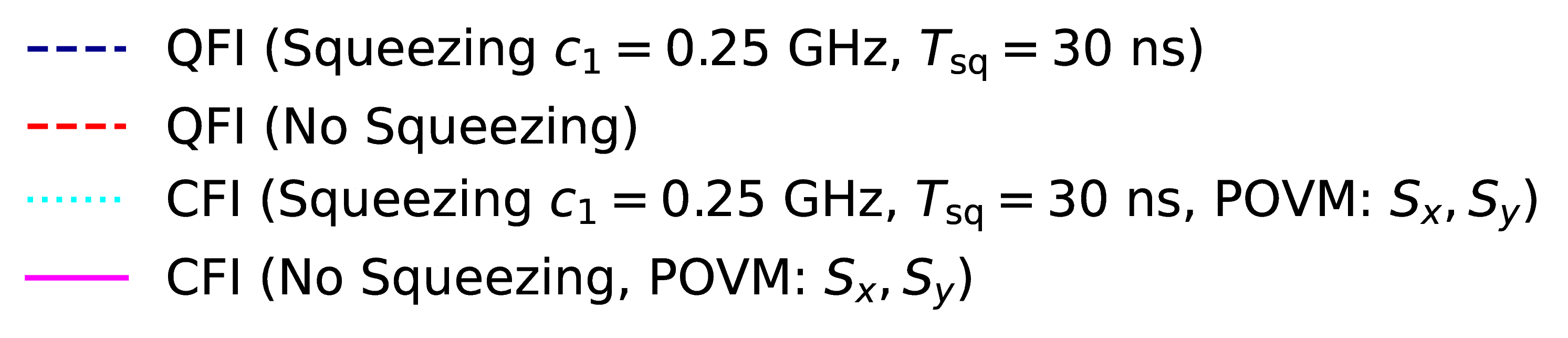}
\caption{QFI and CFI as a function of time for a two-NV-center spin-1 system. The plots account for thermalization and dephasing rates of \( \gamma_\mathrm{t} = 0.2 \) MHz and \( \gamma_\mathrm{d} = 0.1 \, \gamma_\mathrm{t} \). Curve identification, including specific squeezing parameters and POVMs, is provided in the accompanying legend figure.}
\label{FigB1}
\end{figure}

\begin{equation}
\textrm{CFI}(\theta) = \sum_i \frac{1}{p_i(\theta)} \left( \frac{\partial p_i(\theta)}{\partial \theta} \right)^2,
\end{equation}
where $p_i(\theta)$ represents the probability of the $i$-th measurement outcome for a given $\theta$. In our context, these probabilities correspond to outcomes measured along the eigenvectors of $S_x$ and $S_y$, providing the necessary basis for CFI calculation.
By choosing eigenvectors of $S_x$ and $S_y$ as the Positive Operator-Valued Measure (POVM) elements, we model measurements that are feasible in practical setups. Such measurements can be carried out using state tomography, as detailed in \cite{paris2004quantum}.
However, we note that in standard NV center Ramsey experiments, a second \(\pi/2\)-pulse is typically applied at the end of the free evolution time to rotate the phase accumulated in the transverse plane (\(S_x, S_y\)) onto the \(S_z\)-axis, where the population difference is then measurable. Optical state tomography reconstruction of the density matrix allows for accessing the full POVM set. However, this would be slow and resource-intensive. Hence, nearly optimal \(S_z\) measurements are preferable.
In Fig.~\ref{FigB1}, we observe that the CFI is non-zero and is higher with squeezing, which further supports our results.

\bibliographystyle{iopart-num}  
\bibliography{references}

\end{document}